\documentclass[runningheads,fleqn]{svmult}
\usepackage{makeidx}   
\usepackage{graphicx}  
\usepackage{subeqnar}  
\usepackage{multicol}  
\usepackage{taphys}    
\makeindex             
\newcommand{\greeksym}[1]{{\usefont{U}{psy}{m}{n}#1}}
\newcommand{\umu}{\mbox{\greeksym{m}}}

\usepackage{amsmath}   
\mathindent\parindent 
\usepackage{epsfig}

\newcommand{\imai}{\mathrm{i}}
\renewcommand{\d}{{\rm d}}

\begin{document}

\toctitle{Transport in nanostructures: A comparison between 
nonequilibrium Green functions and density matrices}
\title*{
{\rm \footnotesize  to appear in {\em Advances in Solid State
 Physics}, ed. by B. Kramer (Springer 2001)}\\
Transport in nanostructures:\\ A comparison between 
nonequilibrium Green functions and density matrices
}
%
%
\titlerunning{Transport in nanostructures}
%
\author{Andreas Wacker}
\authorrunning{Andreas Wacker}
%
\institute{Institut f{\"u}r Theoretische Physik, 
Technische Universit{\"a}t Berlin,\\
Hardenbergstr. 36, 10623 Berlin, Germany}

\maketitle              

\begin{abstract}
Stationary electric transport in semiconductor nanostructures is studied
by the method of nonequilibrium Green functions. 
\index{nonequilibrium Green functions}
In  the case of sequential tunneling the results are compared
with \index{density matrix theory} density matrix theory, 
providing almost identical
results. Nevertheless, the method of Green functions 
is easier to handle due to the availability of an absolute 
energy scale. It is demonstrated, that the transport
in complicated structures, like quantum cascade lasers,
can be described in reasonable agreement with  experiment.
\end{abstract}

\section{Introduction}
Transport through semiconductor nanostructures\cite{DAT95,FER97,DIT98}
is dominated by quantum effects.
A corresponding \index{quantum transport}
quantum transport theory can be based
on density matrices or nonequilibrium Green functions
(see, e.g., Refs.~\cite{HAU96,KUH98} and references cited therein).
In this article, I want to  show that nonequilibrium Green functions
provide a strong tool to handle this situation for {\em stationary}
transport, even for complicated structures like
the quantum cascade laser. By a direct
comparison with the density matrix method
the differences between both methods are examined
for the test case of sequential tunneling between neighboring quantum wells.

The article is organized as follows:
After a general formulation of the problem (Section \ref{SecGeneral})
it will be shown in Section \ref{SecMethod} 
how  the method of nonequilibrium Green functions
can be applied in the stationary state.
A simple example is discussed in detail in Section
\ref{SecSimple},
and a direct comparison with density matrix theory
is made in Section \ref{SecDMT}.
Finally, the full power of the Green function approach is demonstrated
in a simulation of quantum cascade lasers in Section \ref{SecQCL}.

\section{General aspects of quantum transport\label{SecGeneral}}

The starting point  for a quantum kinetic description is the
Hamilton operator in second 
quantisation (using  basis states labeled by $\alpha$) 
\begin{equation} 
\hat{H}=\hat{H}_0+\hat{U}+\hat{H}_{\rm scatt} 
\end{equation} 
where (in Heisenberg representation)
\begin{equation} 
\hat{H}_0=\sum_{\alpha}E_{\alpha}a^{\dag}_{\alpha}(t)a_{\alpha}(t) 
\label{EqH0}
\end{equation} 
is diagonal in the basis $|\alpha\rangle$, 
\begin{equation} 
\hat{U}=\sum_{\alpha,\beta}U_{\alpha,\beta}(t) 
a^{\dag}_{\alpha}(t)a_{\beta}(t) 
\label{EqHU}
\end{equation} 
describes nondiagonal parts of the Hamiltonian as well as
the presence of electric fields,  and 
$\hat{H}_{\rm scatt}$ refers to interactions with phonons, random 
impurity potentials (which are treated within impurity averaging), 
or interactions between the particles. 
The final goal is to calculate various observables such as
the occupation of the state $\alpha$
\begin{equation} 
f_{\alpha}(t)=\langle a^{\dag}_{\alpha}(t)a_{\alpha}(t)\rangle 
\label{EqDensity}
\end{equation} 
or transition rates
\begin{equation} 
j^{\beta\to\alpha}(t)=\frac{2}{\hbar} 
\Re\left\{\imai U_{\beta,\alpha} 
\langle a_{\beta}^{\dag}(t)a_{\alpha}(t)\rangle\right\} 
\label{EqJdmt} 
\end{equation}
between the respective states, which can be obtained from the
equation of continuity for the occupations.
Here $\langle\ldots  \rangle$ denotes the quantum mechanical expectation value with
the (nonequilibrium) distribution.
Besides these one-particle density matrices 
$\langle a_{\beta}^{\dag}(t)a_{\alpha}(t)\rangle$, higher order
density matrices describe correlation effects and response functions.

Two formalisms exist to treat the quantum problem [i.e. to find approximations in order to
to obtain solutions for Eqs.~(\ref{EqDensity},\ref{EqJdmt})]:
Within the method of density matrices, the temporal evolution of these quantities
(where all operators are taken at the same time)
is studied directly. This method was extremely successful in the study of
electron kinetics on short time scales, see, e.g., Ref.~\cite{KUH98} for details.
On the other hand, Green functions
depend on two different times. These have been used for 
stationary transport and  for electron kinetics, see, e.g., 
Refs.~\cite{HAU96,BON00}  for details.

\section{Method of Nonequilibrium Green Functions\label{SecMethod}}

The key quantities in the theory of nonequilibrium Green functions
are  the {\em correlation function} (or `lesser' Green function) 
\begin{equation} 
G^<_{\alpha_1,\alpha_2}(t_1,t_2)= 
\imai\langle a^{\dag}_{\alpha_2}(t_2)a_{\alpha_1}(t_1)\rangle 
\end{equation} 
which describes the occupation of the states (for equal times and indices), together 
with the respective correlations both in time and state index, 
as well as  the {\em retarded  Green function} 
\begin{equation} 
G^{\rm ret}_{\alpha_1,\alpha_2}(t_1,t_2)=
-\imai\Theta(t_1-t_2) 
\langle a_{\alpha_1}(t_1)a^{\dag}_{\alpha_2}(t_2)
+a^{\dag}_{\alpha_2}(t_2)a_{\alpha_1}(t_1)
\rangle
\end{equation} 
which describes the response of the system at time 
$t_1$ in state $\alpha_1$ after an  excitation at time 
$t_2$ in state $\alpha_2$. 

If the external potential is static and transients resulting from
initial conditions have disappeared, the system is typically in a
stationary state\footnote{This is not neccessarily the case as self-sustained
oscillations or even chaotic behavior  frequently occur in
semiconductor systems at high electric fields \cite{SCH01}.}
and all functions depend only on the time difference $t_1-t_2$. Then it is 
convenient to work in Fourier space defined by 
\begin{equation}
F_{\alpha_1,\alpha_2}(E)=\frac{1}{\hbar}\int \d t\, 
e^{\imai Et/\hbar} F_{\alpha_1,\alpha_2}(t+t_2,t_2)\, .
\end{equation} 
This defines the energy $E$ which is {\em not}  the level  energy $E_{\alpha}$ of a certain
state, but a new parameter, setting an absolute scale to compare energies 
of different states.
Then the following equations determine the Green functions \cite{DAT95,HAU96,MAH90}:
\begin{align} 
\begin{split}
&\left(E-E_{\alpha_1}\right) 
G^{\rm ret/adv}_{\alpha_1,\alpha_2}(E) 
-\sum_{\beta} U_{\alpha_1,\beta}G^{\rm ret/adv}_{\beta,\alpha_2}(E) 
=\\
&\phantom{\left(E-E_{\alpha_1}\right) G^{\rm ret/adv}_{\alpha_1,\alpha_2}(E)-}
\delta_{\alpha_1,\alpha_2}+ 
\sum_{\beta}\Sigma^{\rm ret/adv}_{\alpha_1,\beta}(E) 
G^{\rm ret/adv}_{\beta,\alpha_2}(E) 
\label{EqGretE}  
\end{split}\\
&G^<_{\alpha_1,\alpha_2}(E)=\sum_{\beta,\beta'} 
G^{\rm ret}_{\alpha_1,\beta}(E)\Sigma^{<}_{\beta,\beta'}(E) 
G^{\rm adv}_{\beta',\alpha_2}(E)\label{EqKeldysh}
\end{align} 
where the advanced Green function is just given by 
$G^{\rm adv}_{\beta,\alpha}(E)=\left\{G^{\rm ret}_{\alpha,\beta}(E)\right\}^*$.
These equations contain the self energies, which are typically  functionals 
of the Green functions, and depend on the approximation chosen. To provide a   
glimpse of their structure, two examples are given here:

In the simple Born approximation, the retarded self energy for impurity scattering 
with scattering matrix elements $V_{\alpha_1,\beta}$
is given by:
\begin{equation} 
\Sigma^{\rm ret}_{\alpha_1,\alpha_2}(E)= 
\sum_{\beta} 
\langle V_{\alpha_1,\beta}
V_{\beta,\alpha_2}\rangle_{\rm imp} 
\frac{1}{E-E_{\beta}+\imai 0^+}
\label{EqSigimp} 
\end{equation} 
where $\langle V_{\alpha_1,\beta}
V_{\beta,\alpha_2}\rangle_{\rm imp}$ denotes the averaging over all
impurity configurations.
A comparison with Fermi's golden rule gives
$\Im\{\Sigma^{\rm ret}_{\alpha,\alpha}(E_{\alpha})\}= 
-\hbar/2\tau_{\alpha}$, which relates the imaginary part of
$\Sigma^{\rm ret}$ to the lifetime $\tau_{\alpha}$ of the state.

In the self-consistent  Born approximation, the lesser self-energy for
phonon scattering reads
\begin{equation} 
\begin{split} 
\Sigma^{<}_{\alpha_1,\alpha_2}(E)
=&\sum_{\vec{p},l}\sum_{\beta_1,\beta_2} 
M^{\rm phon}_{\alpha_1,\beta_1}(\vec{p},l)M^{\rm phon}_{\beta_2,\alpha_2}(\vec{p},l)\\
&\times\Big[n_B(\hbar\omega_l(\vec{p}))G^{<}_{\beta_1,\beta_2} 
(E-\hbar\omega_l(\vec{p}))\\
&\phantom{\times\Big[}+\left[n_B(\hbar\omega_l(\vec{p}))+1\right]G^{<}_{\beta_1,\beta_2} 
(E+\hbar\omega_l(\vec{p}))\Big]\, , 
\label{EqSiglessphon} 
\end{split} 
\end{equation} 
where $M^{\rm phon}_{\alpha_1,\beta_1}(\vec{p},l)$ is the matrix element 
of the electron-phonon interaction.
$\vec{p}$ and $l$ denote 
the  wave vector and the mode of the phonon spectrum with corresponding frequency 
$\omega_l(\vec{p})$.
$n_B(E)=(\exp(E/k_BT)-1)^{-1}$ is the Bose distribution, describing the occupation of
phonon modes in thermal equilibrium.
Taking into account that $ G^{<}$ describes the occupation of states,
one can identify the first term as phonon absorption from states with energy
$E-\hbar\omega_l(\vec{p})$, and the second term as the combination of
stimulated and spontaneous phonon emission from states with energy
$E+\hbar\omega_l(\vec{p})$. This demonstrates that 
$\Sigma^{<}_{\alpha_1,\alpha_2}(E)$ is just the in-scattering rate at energy
$E$. Furthermore it becomes clear, that $E$ is the relevant energy scale
for the occupation, and indeed in thermal equilibrium
one finds
\begin{equation} 
G^{<}_{\alpha,\alpha}(E)
=iA_{\alpha}(E)n_F(E-\mu)
\label{EqEquib}
\end{equation} 
with the spectral function 
$A_{\alpha}(E)=-2\Im\{G^{\rm ret}_{\alpha,\alpha}(E)\}$
and the electro-chemical potential $\mu$. 
It is crucial to note, that the energy $E$ enters the Fermi function $n_F(E-\mu)$
instead of  the level energy $E_{\alpha}$.

Eqs.~(\ref{EqGretE},\ref{EqKeldysh}) have to be solved self-consistently
with  the functionals for the self-energies, which is a formidable task.
Finally, the physical  observables of interest can be calculated  via the density matrices
\begin{equation} 
\langle a_{\beta}^{\dag}(t)a_{\alpha}(t)\rangle=
-\imai\int\frac{\d E}{2\pi} 
G^{<}_{\alpha,\beta}(E)\label{EqRho}\, .
\end{equation}
Such calculations have been performed for
the double barrier resonant tunneling diode \cite{LAK97}
and \index{superlattice} superlattice structures \cite{WAC99a,WAC01}.
In Section \ref{SecQCL}, simulation results for a quantum cascade laser 
structure are presented.
In order to study the general concept, an analytic solution
for a simple example will be given in the subsequent section. 

\section{A simple example\label{SecSimple}}
\begin{figure}[tb]
\sidecaption
\includegraphics[width=.7\textwidth]{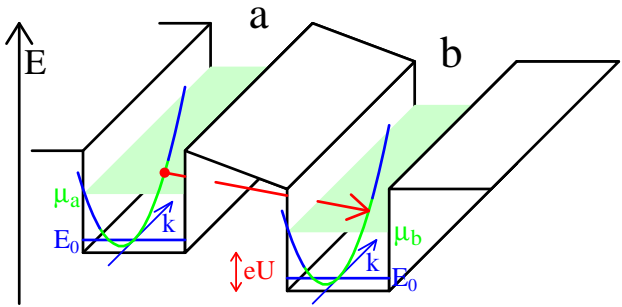}
\caption[ ]{Sketch of the test structure consisting of two 
neighboring quantum wells.
A $k$-conserving transition is indicated}
\label{FigClassic}
\end{figure}
Let us consider sequential tunneling between
two neighboring quantum wells, labeled by $a$ and $b$, see Fig.~\ref{FigClassic}. 
In both wells only the lowest bound state (with energy $E_0$) plays a role and 
the in-plane behavior is determined by plane waves with quasimomentum 
$k=(k_x,k_y)$ and the
dispersion $E_k=\hbar^2k^2/2m$.
Such an experiment has been performed by applying separate
contacts to the individual layers \cite{MUR95,TUR96}.
Reflecting the experimental situation, the following assumptions
are used in the theory: 
\begin{itemize}
\item  The translational invariance of the structure 
allows only for $k$-conserving transitions  $a\to b$
with a tunneling matrix--element $M$.
\item Due to weak coupling between the layers, only a small current
is present, so that both layers are in thermal equilibrium with the
respective contact. 
\item The electron densities in both layers are equal, so that the bias
$U$ determines both the difference in electro-chemical potentials
$\mu_a-\mu_b=eU$, and the shift of the energy level  $b$ by $-eU$.
\item Impurity scattering is the dominant scattering process, which is
realistic for the case of vanishing temperature, $T\to 0$, considered here. 
Correlations in the impurity potential between both wells are neglected.
\end{itemize}
First note, that the simple application of Fermi's golden rule
forbids transitions between the wells (see the arrow in Fig.~\ref{FigClassic})
for $U\neq 0$, due to the
energy conserving $\delta$-function. Therefore a quantum transport
calculation is indispensable.
From the assumptions mentioned above we obtain the Hamiltonian
\begin{equation}
\begin{split}
\hat{H}=&\sum_k (E_0+E_k) a^{\dag}_ka_k+(E_0+E_k -eU)b^{\dag}_kb_k
+Mb^{\dag}_ka_k
+Ma^{\dag}_kb_k\\
&+\sum_{k,k'} V^a_{k,k'}a^{\dag}_ka_{k'}+V^b_{k,k'}b^{\dag}_kb_{k'}
\end{split}
\end{equation}
where the matrix elements $ V^{a/b}_{k,k'}$ for impurity scattering 
are subject to impurity averaging during the calculation.

Now we apply the Green function approach discussed in the last section.
The general state index $\alpha$ is replaced by the two indices $a/b,k$.
Neglecting correlations in the impurity potential between both wells
implies $\langle  V^{\mu}_{k,k'} V^{\nu}_{k',k}\rangle_{\rm imp}\sim \delta_{\mu,\nu}$. 
Therefore the self-energies are diagonal in the
well index and Eq.~(\ref{EqGretE}) takes the form
\begin{equation} 
\begin{split}
&\begin{pmatrix} G_{aa}^{\rm ret}(k;E) & G_{ab}^{\rm ret}(k;E)\\
G_{ba}^{\rm ret}(k;E) & G_{bb}^{\rm ret}(k;E)\end{pmatrix}\\
&\quad =\begin{pmatrix} E-E_0-E_k -\Sigma^{\rm ret}_{aa}(k;E)& -M\\ 
-M & E-E_0-E_k+eU-\Sigma^{\rm ret}_{bb}(k;E)\end{pmatrix}^{-1}
\end{split}
\end{equation}
which gives ${\bf G}^{\rm ret}(k;E)={\bf G}^{\rm ret (0)}(k;E)+M{\bf G}^{\rm ret (1)}(k;E)+{\cal O}(M^2)$ with
\begin{equation}
G^{\rm ret (0)}_{\mu\nu}(k;E)=
\delta_{\mu\nu}\frac{1}{E-E_0-E_k +eU\delta_{\nu,b}-\Sigma^{\rm ret}_{\nu\nu}(k;E)}
\end{equation}
and
\begin{equation}
{\bf G}^{\rm ret (1)}(k;E)=
\begin{pmatrix} 0 & G_{aa}^{\rm ret (0)}(k;E)G_{bb}^{\rm ret (0)}(k;E) \\
G_{bb}^{\rm ret (0)}(k;E)G_{aa}^{\rm ret (0)}(k;E) & 0
\end{pmatrix} 
\end{equation}
Then Eq.~(\ref{EqKeldysh}) yields
\begin{equation}
\begin{split}
G^{<}_{ba}(k;E)=&G_{bb}^{\rm ret (0)}(k;E)MG^{<\, (0)}_{aa}(k;E)+
G^{<\, (0)}_{bb}(k;E)MG_{aa}^{\rm adv (0)}(k;E)\\
&+{\cal O}(M^2)
\end{split}
\end{equation}
The zeroth order terms refer to the uncoupled layers in equilibrium.
Therefore Eq.~(\ref{EqEquib}) applies and with Eqs.~(\ref{EqJdmt},\ref{EqRho})
one obains the total current \cite{MAH90}
\begin{equation}
\begin{split}
I^{a\to b}=&
2\mbox{(for Spin)}\frac{eM^2}{\hbar}\sum_k \int \frac{{\rm d}E}{2\pi}
A_a(k;E)A_b(k;E)\\
&\times \left[n_F(E-\mu_a)-n_F(E-\mu_b)\right]+{\cal O}(M^3)\, .
\end{split}
\end{equation}
\begin{figure}[tb]
\sidecaption
\includegraphics[width=.7\textwidth]{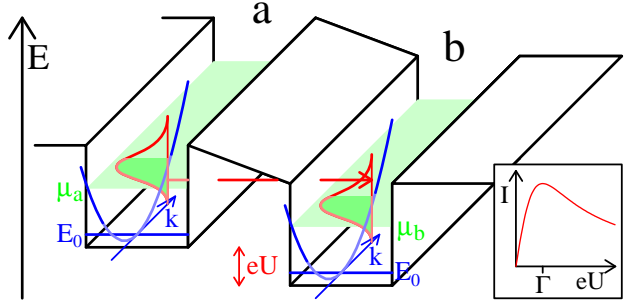}
\caption[ ]{Transitions in the Green-function treatment at fixed
energy $E$. Transitions are allowed if
$A_a(k,E)$ and $A_b(k,E)$ overlap in the energy window between
the electrochemical potentials. The inset depicts the current-voltage characteristic}
\label{FigSpekt}
\end{figure}

This result shows that the transitions occur at fixed energy $E$, see 
Fig.~\ref{FigSpekt}.
The spectral functions $A_a(k,E)$ and $A_b(k,E)$ give the weight
of the levels $a$ and $b$ at the energy $E$, respectively.
For each energy, there is a contribution to the current
if there is a finite weight of both states 
and difference in occupation  $\left[n_F(E-\mu_a)-n_F(E-\mu_b)\right]$.
This demontrates the crucial role of $E$ as an energy scale.
Following Sec. 3.3 of Ref.~\cite{WAC01} this expression can be simplified
for  $\Sigma^{\rm ret}_{\nu\nu}(k;E)=-i\Gamma/2$:
\begin{equation}
\begin{split}
I^{a\to b}_{GFT}
\approx 
A\frac{e|M|^2\rho_0}{\hbar} 
\frac{2\Gamma}{(eU)^2+\Gamma^2}
\int {\rm d}E \left[n_F(E-\mu_a)-n_F(E-\mu_b)\right]
\label{EqJNGFsimp}
\end{split}
\end{equation}
where $\rho_0=m/(\pi\hbar^2)$ is the two-dimensional density of states including spin and $A$
is the sample area. 
Eq.~(\ref{EqJNGFsimp}) was used in the interpretation of the experiments \cite{MUR95,TUR96}
where good agreement was found.

\section{Treatment by density matrices\label{SecDMT}}
For comparison, the example of  the preceding section 
will now be treated within the density matrix approach. 
Here the key quantities are the occupations
$f_a(k)=\langle a^{\dag}_ka_k\rangle$ and 
$f_b(k)=\langle b^{\dag}_kb_k\rangle$ as well as the polarisation $P(k)=\langle b^{\dag}_ka_k\rangle$,
which provides the  transition rates according to Eq.~(\ref{EqJdmt}).
The key idea is to derive the temporal evolution of these quantities as a set
of differential equations and to solve these after some approximations.
Thus
\begin{equation}
\begin{split}
\frac{\hbar}{\imai}\frac{\d}{\d t} P(k,t)=&\langle [\hat{H},b^{\dag}_k(t)a_k(t)]\rangle\\
=&-eU P(k,t)
+M\left[f_a(k,t)-f_b(k,t)\right]\\
&+\sum_{k'} \left[
V_{k'k}^{b} \langle b_{k'}^{\dag}(t)a_k(t)\rangle-
V_{kk'}^{a} \langle b_k^{\dag}(t)a_{k'}(t)\rangle  \right]
\label{EqPdyn}
\end{split}
\end{equation}
The density matrices 
$\langle b_{k}^{\dag}(t)a_{k'}(t)\rangle$
contain the phase of the scattering  matrix-element $V_{kk'}^{a}$. Thus one
evaluates the temporal evolution as
\begin{equation}
\begin{split}
\frac{\hbar}{\imai }\frac{\d}{\d t}
V_{kk'}^{a}\langle b_{k}^{\dag}(t)a_{k'}(t)\rangle
=&(E_{k}-eU-E_{k'})V_{kk'}^{a}\langle b^{\dag}_{k}a_{k'}\rangle\\
+&M\left[V_{kk'}^{a}\langle a_{k}^{\dag}a_{k'}\rangle-
V_{kk'}^{a}\langle b_{k}^{\dag}b_{k'}\rangle\right]
\\
+&\sum_{\bf k''} V_{k''k}^{b}V_{kk'}^{a}
\langle b_{k''}^{\dag}a_{k'}\rangle
-\sum_{\bf k''}V_{kk'}^{a} V_{k'k''}^{a}
\langle b_{k}^{\dag}a_{k''}\rangle
\end{split}
\end{equation}
After impurity averaging one finds
$\langle V_{k''k}^{\nu}V_{kk'}^{a}\rangle_{\rm imp}= |V_{kk'}^{a}|^2 
\delta_{\nu a}\delta_{k'',k'}$ and we have
\begin{equation}
\begin{split}
\frac{\hbar}{\imai }\frac{\d}{\d t}
V_{kk'}^{a}\langle b_{k}^{\dag}(t)a_{k'}(t)\rangle
=&(E_{k}-eU-E_{k'})V_{kk'}^{a}
\langle b^{\dag}_{k}a_{k'}\rangle
- |V_{kk'}^{a}|^2 P(k,t)\\
&+M\left[V_{kk'}^{a}\langle a_{k}^{\dag}a_{k'}\rangle-
V_{kk'}^{a}\langle b_{k}^{\dag}b_{k'}\rangle\right]
\label{EqVba}
\end{split}
\end{equation}
In Eq~(\ref{EqVba}) it is tempting to neglect the last line, as
these terms contain both  $M$ and $V$.
In this case one obtains
\begin{equation}
V_{kk'}^{a}\langle b_{k}^{\dag}(t)a_{k'}(t)\rangle=
\frac{1}{E_{k}-eU-E_{k'}+\imai 0^+}|V_{kk'}^{a}|^2
P(k)
\end{equation}
in the stationary state, where the term 
$\imai 0^+$ ensures that correlations vanish for  $t\to -\infty$.
Therefore
\begin{equation}
\sum_{k'}V_{kk'}^{a} \langle b_k^{\dag}(t)a_{k'}(t)\rangle 
=\Sigma^{\rm ret}_a(k,E_0+E_k-eU)P(k)
\end{equation}
where Eq.~(\ref{EqSigimp}) has been used.
In the same way one finds $\sum_{k'} V_{k'k}^{b} \langle b_{k'}^{\dag}(t)a_k(t)\rangle=
\Sigma^{\rm adv}_b(k,E_0+E_k)P(k)$.
Inserting into Eq.~(\ref{EqPdyn}) gives the polarisation
\begin{equation}
P(k)=\frac{M\left[f_{a}(k)-f_{b}(k)\right]}
{eU+\Sigma^{\rm ret}_a(k,E_0+E_k-eU)-\Sigma^{\rm adv}_b(k,E_0+E_k)}
\end{equation}
and with 
$\Sigma^{\rm adv}_b(k,E_0+E_k)-\Sigma^{\rm ret}_a(k,E_0+E_k-eU)\approx i\Gamma$ 
the current 
\begin{equation}
I^{a\to b}_{\rm DMTsimp}= 2\mbox{(for Spin)}e\frac{M^2}{\hbar}\sum_{k}
\frac{2\Gamma}{(eU)^2+\Gamma^2}
\left[f_{a}(k)-f_{b}(k)\right] \, . \label{EqJDMTsimp}
\end{equation}
This means that the transitions occur between states with
the same $k$ as indicated  in Fig.~\ref{FigClassic}.
In contrast to the treatment by Fermi's golden rule,
life-time broadening allows for these transitions even  for  finite $U$.
This structure appears frequently in density matrix calculations, 
see, e.g., Ref.~\cite{KAZ72} for sequential tunneling.
On first sight $I^{a\to b}_{\rm DMTsimp}$ seems to be similar to $I^{a\to b}_{\rm GFT}$.
But for the situation considered, $I^{a\to b}_{\rm DMTsimp}$ vanishes, because
$f_{a}(k)\equiv f_{b}(k)$ as the electron density is identical in both wells.
This indicates that the result is not trustworthy.

Now we take into account the first term in the
last line of Eq~(\ref{EqVba}), which indeed is of the same order
as the other terms because $P(k)$ is of order $M$. Therefore we have to evaluate
\begin{equation}
\begin{split}
\frac{\hbar}{\imai }\frac{\d}{\d t}
V_{kk'}^{a}\langle a_{k}(t)^{\dag}a_{k'}(t)\rangle
=&(E_k-E_{k'})V_{kk'}^{a}\langle a_{k}^{\dag}a_{k'}\rangle\\
+&M\left[V_{kk'}^{a}\langle b_{k}^{\dag}a_{k'}\rangle-
V_{kk'}^{a}\langle a_{k}^{\dag}b_{k'}\rangle\right]\\
+&\sum_{k''}
V_{k''k}^aV_{kk'}^{a}\langle a_{k''}^{\dag}a_{k'}\rangle
-V_{kk'}^{a}V_{k'k''}^a\langle a_{k}^{\dag}a_{k''}\rangle
\end{split}
\end{equation}
Perfoming impurity averaging and neglecting the $M$ terms 
(yielding expressions of order $M^2$) gives:
\begin{equation}
V_{kk'}^{a}\langle a_{k}(t)^{\dag}a_{k'}(t)\rangle
=\frac{-1}{E_k-E_{k'}+i0^+}
|V_{kk'}^{a}|^2\left[f_a(k')-f_a(k)\right]
\end{equation}
Now the solution of Eq~(\ref{EqVba}) provides
\begin{equation}
\begin{split}
V_{kk'}^{a}\langle b_{k}^{\dag}(t)a_{k'}(t)\rangle=&
\frac{1}{E_{k}-eU-E_{k'}+\imai 0^+}|V_{kk'}^{a}|^2
P(k)\\
+&\frac{M}{eU}
\left[\frac{1}{E_{k}-eU-E_{k'}+\imai 0^+}-\frac{1}{E_k-E_{k'}+i0^+}\right]\\
&\times|V_{kk'}^{a}|^2\left[f_a(k')-f_a(k)\right]\, .
\end{split}
\end{equation}
Using
$\Im\{1/(x+\imai 0^+)\}=-\pi\delta(x)$ and neglecting
the corresponding real part one obtains
\begin{equation}
\begin{split}
\sum_{k'}&V_{kk'}^{a} \langle b_k^{\dag}(t)a_{k'}(t)\rangle 
\approx \Sigma^{\rm ret}_a(k,E_0+E_k-eU)P(k)\\
&+\imai\frac{M}{eU}\Im\{\Sigma^{\rm ret}_a(k,E_0+E_k-eU)\}
\left[n_a(E_k-eU)-n_a(E_k)\right]
\end{split}
\end{equation}
where
$n_a(E_k)=f_a(k)$.
In the same way one obtains
\begin{equation}
\begin{split}
\sum_{k'}& V_{k'k}^{b} \langle b_{k'}^{\dag}(t)a_k(t)\rangle\approx
\Sigma^{\rm adv}_b(k,E_0+E_k)P(k)\\
&+\imai\frac{M}{eU}\Im\{\Sigma^{\rm adv}_b(k,E_0+E_k)\}
\left[n_b(E_k)-n_b(E_k+eU)\right]
\end{split}
\end{equation}
Inserting into Eq.~(\ref{EqPdyn}) and using
$\Sigma^{\rm adv/ret}\approx\pm i\Gamma/2$ one obtains after a
few lines of algebra
\begin{equation}
\begin{split}
I^{a\to b}_{\rm DMT}=&  2e\frac{M^2}{\hbar }\sum_{k}
\frac{2\Gamma}{(eU)^2+\Gamma^2}\\
&\times\left[\frac{n_{a}(E_k)-n_{b}(E_k+eU)}{2}
+\frac{n_{a}(E_k-eU)-n_{b}(E_k)}{2}\right]  
\label{EqJDMTkorr}
\end{split}
\end{equation}
\begin{figure}[tb]
\sidecaption
\includegraphics[width=.7\textwidth]{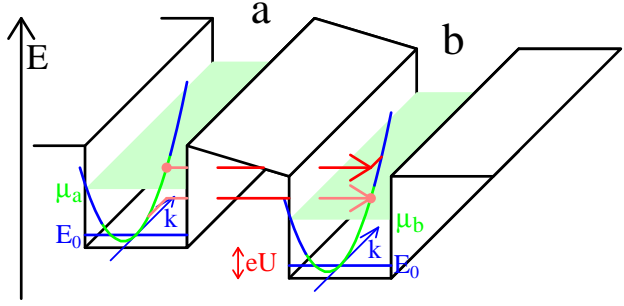}
\caption[ ]{Transitions in the density matrix treatment according
to Eq.~(\ref{EqJDMTkorr}). The current is driven by
impurity assisted transitions between states of different $k$, 
so that the energy is conserved}
\label{FigJDMTKorr}
\end{figure}
This means, that the current is driven by the difference in 
occupation at the same
energy, rather than the difference in occupation at the same $k$.
Fig.~\ref{FigJDMTKorr} suggests that these transitions should be viewed
as scattering-assisted transitions.
Note, that this interpretation is completely
different from the result by Eq.~(\ref{EqJDMTsimp}). 
The fact, that the transitions occur at fixed energy
strongly resembles the Green-function formalism, see
Fig.~\ref{FigSpekt}, and the evaluation of Eq.~(\ref{EqJNGFsimp}) and
Eq.~(\ref{EqJDMTkorr}) give similar results. 
The difficulty in obtaining this result in density matrix theory
seems to be related to the fact that, unlike in the Green
function formalism, the energy scale $E$ is not available.

\section{Results for quantum cascade laser structures \label{SecQCL}}
Electrically driven unipolar semiconductor lasers can be made by a
proper design of nanostructures. The  key point is 
population inversion between subbands, which
can be achieved by a rather complicated 
sequence of quantum wells.
In order to amplify this effect, several
periods of such sequences are grown on top of each
other, resulting in a quantum cascade laser \cite{FAI94}
\index{quantum cascade laser}
operating in the 10 $\umu$m range.

\begin{figure}[tb]
\sidecaption
\includegraphics[width=0.7\textwidth]{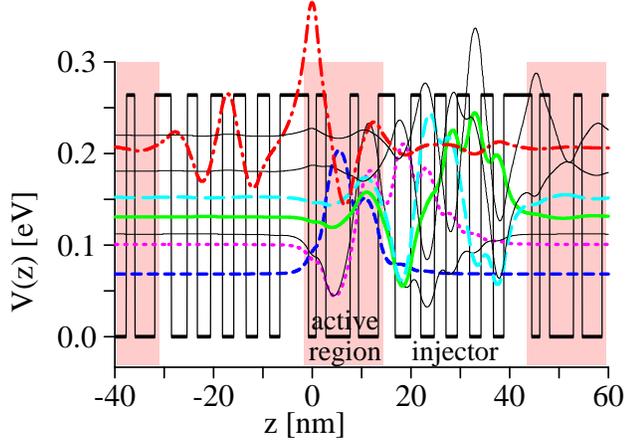}
\caption[ ]{Conduction band of the quantum cascade laser structure
from  Ref.~\cite{SIR98}. The eight lowest Wannier functions  are given 
for one period, where the base line depicts
the respective level energy. Note that these functions 
are repeated in each period}
\label{FigQCLWan}
\end{figure}

In Fig.~\ref{FigQCLWan} such a structure (from  Ref.~\cite{SIR98}) is shown.
Here one period consists of an active region (where the
lasing transition occurs) of 3 wells ($0< z < 15$ nm)
and the injector region of 5 wells (for  $15 {\rm nm} < z < d$).
This sequence with period $d=45.3$ nm is repeated 30 times in the experiment.
 
Similar to the treatment in Section \ref{SecSimple},
we use a set of basis states consisting of localized states in the
growth direction (labeled by  $\nu$) 
and a free particle behavior with wave vector $k$ 
perpendicular to the growth direction.
For the localized states we use  Wannier functions as 
depicted in Fig.~\ref{FigQCLWan}.
This allows for a calculation of all matrix elements in
Eqs.~(\ref{EqH0},\ref{EqHU})
from the sample parameters following
Ref.~\cite{WAC01}. 
For field strengths of the order of $F\approx 50$ kV/cm, the states
in the injector become aligned to the 
excited state (dashed-dotted line) in the active region of the next period.
This design provides a strong enhancement of population
of this level associated with with population inversion in the active
region. 

The self-energies are evaluated within the self-consistent Born
approximation treating scattering with impurities, optic
phonons, and acoustic phonons similar to Ref.~\cite{WAC01}.
Eqs.~(\ref{EqGretE},\ref{EqKeldysh}) were solved
self-consistently with the functionals for the self-energies
reaching an accuracy of approimately 1\%.
Details will be given elsewhere \cite{LEE01}.
For practical purposes the momentum dependence
of the matrix elements had to be neglected. Thus,
these quantum transport calculations can be considered as 
complementary to semiclassical Monte-Carlo 
simulations \cite{TOR99,IOT00}, 
where such details can be easily taken into account.

\begin{figure}[tb]
\sidecaption
\includegraphics[width=.6\textwidth]{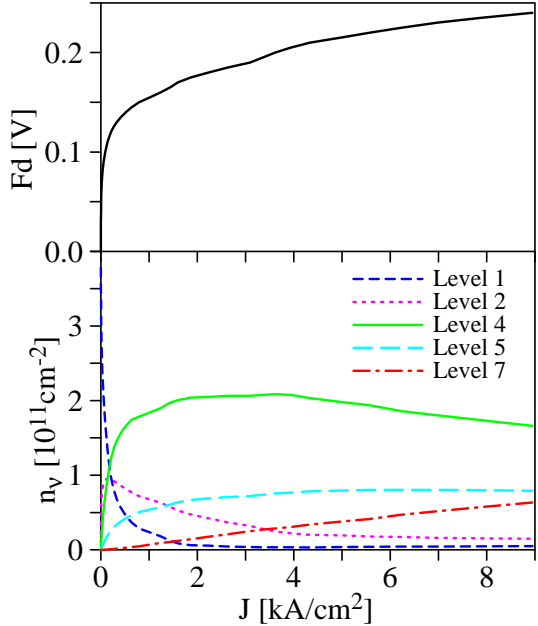}
\caption[ ]{Voltage drop per period $Fd$ 
as well as occupation $n_{\nu}$ for different  levels 
versus current density.
The linestyle corresponds to the fat lines in Fig.~ \ref{FigQCLWan}.
The occupation of the other levels is below $5\times 10^{10}/{\mathrm cm}^2$
for all currents}
\label{FigQCLresult}
\end{figure}

Fig.~\ref{FigQCLresult} shows some results from our calculations.
The field-current relation  is in good agreement with the experimental
findings of  Ref.~\cite{SIR98}. 
The electron densities of level $\nu$ are evaluated by
\begin{equation}
n_{\nu}=\frac{2\mbox{(for spin)}}{A}\sum_k \int \frac{\d E}{2\pi} \Im\{G_{\nu\nu}^<(k;E)\}\, .
\end{equation}
For vanishing current and field, essentially the lowest level is occupied.
With increasing field, the energy of the levels 4, 5, and 6 is diminished because
they are located further to the right. Thus, the carriers are transfered to these
states, which are located in the injector region. At the same time the occupation
of level 7 increases linearly with the current, which indicates that the transport
essentially occurs via this level. (The sample was designed precisely for this behavior.)
At moderate currents, occupation inversion occurs between this level 7
and the levels 1, 2, and 3, which are lower in energy. As the dipole matrix element
between these levels and level 7 is rather large (all of them are essentially
located in the active region), lasing is likely
to occur, which is indeed observed in the device for current densities
above 7 kA/cm$^2$.

\section{Conclusion}

Nonequilibrium Green functions provide a powerful tool to
study quantum transport in nanostructures under stationary
conditions. They allow for a quantitative description
of nanostructure devices such as superlattices and quantum
cascade lasers. The power of this method results from
the availability of the global energy scale $E$, the 
frequency related to the time difference in $G(t_1,t_2)$. 
Although essentially the same results can be obtained
in density matrix theory, the structure of the equations
appear more complicated there.

Fruitful cooperation with S.-C. Lee
and financial support by DFG within SFB296 and FOR394
are gratefully acknowledged.

\addcontentsline{toc}{section}{References}

\end{document}